\documentclass{article}
\usepackage{spconf,amsmath,graphicx,hyperref}
\usepackage{adjustbox}
\usepackage{enumitem}
\usepackage{booktabs}     
\usepackage{array}        
\usepackage{graphicx}     
\usepackage{xcolor}       
\usepackage{caption}      
\usepackage[table]{xcolor}  
\usepackage{multirow}
\usepackage{booktabs}
\usepackage{amssymb}
\usepackage{url}

\begin{document}
\def\x{{\mathbf x}}
\def\L{{\cal L}}

\newcommand{\given}{\,|\,}
\def\rr{\textcolor{red}}
\def\bb{\textcolor{blue}}
\def\re{\textcolor{black}}

\title{RETLLM: TRAINING AND DATA-FREE MLLMS FOR \\MULTIMODAL INFORMATION RETRIEVAL}
%
\name{Dawei Su\quad \quad Dongsheng Wang\thanks{Corresponding author: dongshengwang@szu.edu.cn.\\ Project supported by the Young Scientists Fund of the National Natural Science Foundation of China under Grant No.62506237.}}
\address{College of Computer Science and Software Engineering \\Shenzhen University, Shenzhen, China}
%
%

\ninept
\maketitle
\begin{abstract}
Multimodal information retrieval (MMIR) has gained attention for its flexibility in handling text, images, or mixed queries and candidates. Recent breakthroughs in multimodal large language models (MLLMs) boost MMIR performance by incorporating MLLM knowledge under the contrastive finetuning framework. However, they suffer from pre-training inconsistency and require large datasets. In this work, we introduce a novel framework, \texttt{RetLLM}, designed to query MLLMs for MMIR in a training- and data-free manner. Specifically, we formulate MMIR as a similarity score generation task and prompt MLLMs to directly predict retrieval scores in a coarse-then-fine pipeline. At the coarse stage, a top-k filtering strategy builds a small yet high-quality candidate pool for each query, enabling MLLMs to focus on semantically relevant candidates. Subsequently, the retrieval score is predicted by feeding both the query and candidate into MLLMs at the fine stage. Importantly, we propose a visual enhancement module during reasoning to help MLLMs re-pick forgotten visuals, improving retrieval. Extensive experiments on MMIR benchmarks show that \texttt{RetLLM} outperforms fine-tuned models. Ablation studies further verify each component. Our work demonstrates that MLLMs can achieve strong MMIR performance without any training, highlighting their inherent multimodal reasoning ability in a simple, scalable framework.
We release our code at: \url{https://github.com/alivecat05/RETLLM}

\end{abstract}
\section{Introduction}
\label{sec:intro}

Multimodal information retrieval (MMIR) systems are expected to search across various modalities and provide relevant pieces of information according to user inputs, where queries and candidates can consist of pure images, text, or a composition of both. These systems play a crucial role in various downstream tasks, such as image-text retrieval, visual question answering (VQA), and retrieval-augmented generation (RAG)~\cite{lewis2020retrieval}. As a pioneering algorithm in multimodal
representation learning, CLIP~\cite{radford2021learning} has demonstrated strong performance in image-text retrieval via aligning each modality into a shared embedding space with contrastive training on image-text
pairs. However, due to its reliance on modality-specific encoders, CLIP fails to cover more challenging cases, such as long-form text and interleaved image-text content.

In parallel, recent studies explore using multimodal large language models (MLLMs) as universal encoders by replacing CLIP-style embeddings with MLLM-derived representations~\cite{wang2023tuning}\cite{wei2024uniir}\cite{jiang2024e5}, treating MLLMs as universal encoders by inserting a summarization prompt: “\textit{\textless query\textgreater. Summarize above sentences in one word:}”, where \textit{\textless query\textgreater} denotes the multimodal content. Then, a contrastive loss applied to the last token is used to fine-tune the parameters of MLLMs. For example, E5-V~\cite{jiang2024e5} trains MLLMs on text pairs using unimodal contrastive learning, achieving strong performance on complex multimodal retrieval tasks. Follow-up works~\cite{lin2024mm}\cite{lan2025llave}\cite{liu2025lamra}\cite{gu2025breaking}enhance E5-V via large-scale multimodal data, two-stage training, and hard negative mining. In addition to embedding learning, recent studies~\cite{chen2024mllm} treat MLLMs as rerankers to refine retrieval results, often requiring specialized training strategies such as noise injection. Despite these advances in MMIR performance, training-based approaches still exhibit several limitations:\textit{1) Objective misalignment: }the inconsistency between autoregressive pretraining and contrastive fine-tuning may undermine the multimodal reasoning capabilities of MLLMs; and \textit{2) Scalability bottleneck:} the dependence on massive multimodal training pairs requires expensive collection costs and computational resources, limiting practical applications.

To address the above shortcomings, we introduce \texttt{RetLLM}, aiming at exploring the zero-shot retrieval potential of MLLMs in a training and data-free manner. Inspired by the recent success of MLLMs in regression tasks via string-based numeric prediction~\cite{lin2024mm}, \texttt{RetLLM} reformulates retrieval as a similarity score prediction task, enabling complex queries such as long-text or compositional inputs without fine-tuning. To balance efficiency and accuracy, \texttt{RetLLM} completes the retrieval in a coarse-then-fine pipeline. Given a  multimodal query, we first collect a small-sized and high-quality candidate pool by employing a lightweight embedding-based model (e.g.,CLIP similarity). This coarse selection filters out samples with low semantic relevance to the query, not only reducing the retrieval time but also allowing MLLMs to focus more on the hard candidates. During the fine-selection stage, we prompt the MLLMs to predict the similarity score of the query and each candidate by feeding both of them into a multimodal instruction. The final prediction is obtained by choosing the candidate with the largest semantic score. 

Importantly, recent studies have shown that hallucinations in MLLMs often lead to impractical responses~\cite{zou2024look}. To address this, a visual enhancement strategy is developed to view visual tokens as supplementary evidence and allow MLLMs to re-pick up the forgotten features during the prediction process. Lastly, we further design an entropy-based decision-making strategy to deal with tied cases where multiple candidates receive the same highest similarity score in the fine stage. This enables our \texttt{RetLLM} to consider the uncertainty score among confusing candidates, resulting in higher retrieval results. 

\begin{figure*}[!th]
\centering
\begin{adjustbox}{width=0.94\textwidth}
\includegraphics{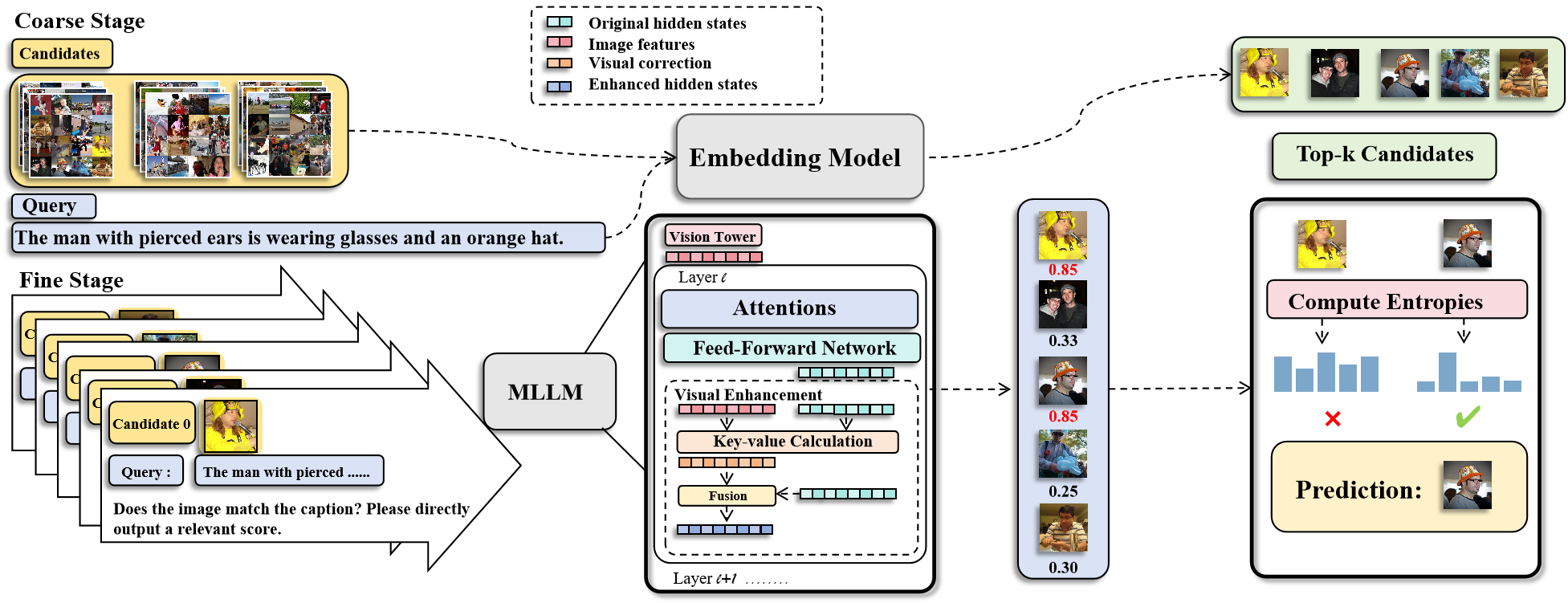}
\end{adjustbox}
\caption{Overview of the \textbf{RetLLM} framework, which integrates Top-\textit{k} filtering, vision enhancement, and entropy-based selection for effective multimodal retrieval}
\label{framework}
\end{figure*}

We summarize our contributions as follows:
\begin{itemize}
    \item We reformulate the multimodal retrieval task as a similarity score generation task, and show that MLLMs possess strong potential for various discriminative tasks.
    \item We introduce \texttt{RetLLM}, a training and data-free framework designed to employ MLLMs for MMIR, where the coarse-then-fine strategy is adopted for quick yet precise retrieval.
    \item Extensive experiments on image-text retrieval and composed image retrieval tasks demonstrate the effectiveness of \texttt{RetLLM}. It outperforms CLIP-based baselines while achieving performance comparable to training-based MLLM models.
\end{itemize}

\section{Methodology}
\label{sec:Methodology}

We define a user query $q$ and $N$ candidates $\Omega = \{c_1, c_2, ...,c_N\}$, where each $q$ and $c_n$ can be an image, text, or interleaved image-text content. In this work, we focus on top-1 retrieval accuracy and aim to search for the best target $c$ for each $q$. Our idea is very simple: we employ an MLLM as a similarity score generator and prompt the pre-trained MLLMs to generate the retrieval score of $q$ and $c_n$ by feeding them into the input instruction. As shown in Fig.~\ref{framework}, our \texttt{RetLLM} performs MMIR in a coarse-then-fine retrieval framework to balance the efficiency and accuracy. A visual enhancement and entropy-based decision-making are further developed to improve the final retrieval performance.

\subsection{Coarse-Then-Fine Framework}
\textbf{Coarse Selection via Semantic Similarity.}
Intuitively, one can directly prompt MLLMs to generate the similarity score between the query $q$ and each candidate $c \in \Omega$. Unfortunately, this naive attempt requires $N$ MLLM queries, leading to heavy time consumption. Generally, only a few samples in $\Omega$ act as valuable candidates of $q$, we thus introduce the coarse-selection module, which forms a small-sized high-quality candidate pool $\mathcal{C}$ for each $q$. Mathematically, we select valuable candidates according to their semantic similarity to $q$:
\begin{equation}
    \mathcal{C} = \text{TopK}(s), \quad s_i=\frac{\mathbf{q}^\top \mathbf{c}_i}{||\mathbf{q}|| ||\mathbf{c}||}, \quad i = 1, 2, \dots, N,
\end{equation}
where $\mathbf{q}$ and $\mathbf{c}$ denote the features of the query and candidate, respectively. $\text{TopK}$ denotes that we select $k$ candidates with the largest similarity scores $s$ from $\Omega$ to form candidate pool $\mathcal{C}$, serving as input for the subsequent fine-grained reranking stage.

\noindent\textbf{Fine-grained Selection with MLLMs.} As discussed above, the candidate pool $\mathcal{C}$ contains hard samples that show high semantic relevance to $q$, and embedding-based models (such as CLIP) fail to distinguish them correctly. Motivated by the impressive reasoning and generation ability of recent MLLMs, we view the retrieval task as a similarity score prediction problem. Specifically, unlike previous models that calculate the retrieval score in the embedding space, we here expect the MLLMs to generate it directly:
\begin{equation}\label{eq2}
    f_i = \text{MLLM} (q,c_i), \quad c_i \in \mathcal{C}.
\end{equation}
where an instruction template (shown in Fig.~\ref{framework}) is designed to take the query $q$ and its candidate $c$ as inputs and prompt MLLMs to predict the semantic similarity scores between them. Due to the small size of $\mathcal{C}$, this fine-grained selection process reduces the MLLM query time from $N$ to $K$, allowing the MLLMs to focus more on the hard candidates

Note that our coarse-then-fine selection framework can be viewed as a hybrid algorithm that combines the representation learning of embedding-based models and the multimodal reasoning capabilities of MLLMs. \texttt{RetLLM} first retrieves $K$ high-quality candidates according to the semantic features in the embedding space, and then explores the pre-trained knowledge encoded in MLLMs to understand the fine-grained differences between the query and hard candidates. The former stage offers fast inference speed but with inaccurate values, while the latter contributes fine-grained similarity scores but with low inference speed. The proposed framework combines the strengths of both under the coarse-then-fine strategy, effectively balancing efficiency and accuracy.

\begin{table*}[!t]
\centering
\small
\setlength{\tabcolsep}{6pt}
\renewcommand{\arraystretch}{0.8}
\begin{tabular}{@{}lccccccccccc@{}}
\toprule
& \multicolumn{4}{c}{\textbf{Short Caption Retrieval}} 
& \multicolumn{4}{c}{\textbf{Long Caption Retrieval}} 
& \multicolumn{3}{c}{\textbf{Compositional Retrieval}} \\
\cmidrule(lr){2-5} \cmidrule(lr){6-9} \cmidrule(l){10-12}
\textbf{Method} 
& \multicolumn{2}{c}{\textbf{Flickr30K}} 
& \multicolumn{2}{c}{\textbf{COCO}} 
& \multicolumn{2}{c}{\textbf{ShareGPT4V}} 
& \multicolumn{2}{c}{\textbf{Urban1K}} 
& \textbf{Replace} & \textbf{Swap} & \textbf{Add} \\ 

\cmidrule(lr){2-3} \cmidrule(lr){4-5} \cmidrule(lr){6-7} \cmidrule(lr){8-9} \cmidrule(l){10-12}
& $q^i \rightarrow c^t$ & $q^t \rightarrow c^i$ 
& $q^i \rightarrow c^t$ & $q^t \rightarrow c^i$ 
& $q^i \rightarrow c^t$ & $q^t \rightarrow c^i$ 
& $q^i \rightarrow c^t$ & $q^t \rightarrow c^i$ 
& & & \\
\midrule
CLIP(ViT-L)
& 87.2 & 67.3 & 58.1 & 37.0 & 81.8 & 84.0 & 47.0 & 47.0 & 79.5 & 62.7 & 74.9 \\
EVA-CLIP 
& 93.9 & 78.8 & 68.8 & 51.1 & 93.1 & 81.2 & 80.4 & 77.8 & 85.9 & 70.3 & 86.7 \\
E5V
& 88.7 & 79.5 & 62.0 & 52.0 & 85.1 & 82.1 & 88.9 & 83.2 & 86.3 & 67.6 & 66.9 \\
VLM2Vec
& 90.6 & 76.0 & 66.6 & 46.8 & 89.8 & 86.9 & 91.3 & 82.4 & 89.5 & 64.8 & 94.2 \\
UniME
& 93.4 & 81.9 & 70.1 & 53.7 & 97.2 & 93.9 & \textbf{95.9} & \textbf{95.2} & 89.0 & 71.6 & 94.4 \\
\midrule
\rowcolor{gray!20}
\texttt{RetLLM}
& \textbf{94.5} & \textbf{82.0} & \textbf{70.4} & \textbf{54.1} 
& \textbf{97.6} & \textbf{94.2} & 88.9 & 78.6 
& \textbf{94.8} & \textbf{92.7} & \textbf{96.2} \\
\bottomrule
\end{tabular}
\label{tab:diverse}
\caption{Results of \textbf{zero-shot} text-image retrieval on short caption datasets (Flickr30K and MS-COCO), long caption datasets
(ShareGPT4V and Urban1K) and compositional benchmark (SugarCrepe). The best results are shown in bold.}
\label{tab:diverse}
\end{table*}

\subsection{Visual Enhancement \& Entropy-based Decision Making} 
Previous work~\cite{wang2024instruction}\cite{rohrbach2018object} shows that due to fine-grained modality imbalance, MLLMs often suffer from hallucinations by losing fine-grained visual details during generation. Inspired by previous works for addressing hallucinations~\cite{zou2024look,lin2025multi,wang2024mllm}, we perform visual re-injection within the Feed-Forward Network (FFN) of the Transformer blocks. Specifically, we first reformulate the standard FFN as a key-value retrieval process. Let $\mathbf{x} \in \mathbb{R}^d$ be the input hidden state of the FFN, with its vanilla form defined as:
\begin{equation}
\mathrm{FFN}(\mathbf{x}) = \phi(\mathbf{x} \mathbf{W_1}) \mathbf{W_2}^\top,
\end{equation}
where $\phi$ is the activation function (e.g., ReLU or SiLU), and $\mathbf{W_1}, \mathbf{W_2} \in \mathbb{R}^{d \times D}$ are the weight matrices (typically $D = 4d$). We can rewrite $\mathbf{W_1}$ and $\mathbf{W_2}$ as:
\begin{equation}
\mathbf{W_1} = (\mathbf{k_1}, \mathbf{k_2}, \dots, \mathbf{k_D}), \quad \mathbf{W_2} = (\mathbf{v_1}, \mathbf{v_2}, \dots, \mathbf{v_D}),
\end{equation}
where $k_i, v_i \in \mathbb{R}^d$ denote the $i$-th key and value vectors, respectively. As a result, the FFN can be reformulated as
\begin{equation}
\mathrm{FFN}(\mathbf{x}) = \sum_{i=1}^{D} \phi(\langle \mathbf{x}, \mathbf{k_i} \rangle) \cdot \mathbf{v_i}.
\end{equation}
This formulation reveals that the FFN acts as a memory module, using $x$ as a query to retrieve the relevant values. Intuitively, we introduce the visual token set $Z_v = \{z_{v,1}, \dots, z_{v,N_v}\}$ as supplementary ``visual knowledge". When activating visual re-inject in layer $l$, we treat visual tokens as new key-value entries and compute the correction term:
\begin{equation}
\Delta(\mathbf{x} \propto \mathbf{Z_v}) = \sum_{j=1}^{N_v} \phi(\langle \mathbf{x}, \mathbf{z_{v,j}} \rangle) \cdot \mathbf{z_{v,j}}.
\end{equation}
Finally, the vanilla FFN output is fused with the visual correction:
\begin{equation}
\mathrm{FFN}^{(l)}(\mathbf{x} \propto \mathbf{Z_v}) = \alpha \Delta(\mathbf{x} \propto \mathbf{Z_v}) + (1 - \alpha) \mathrm{FFN}(\mathbf{x}),
\end{equation}
where $\alpha \in [0,1]$ is the injection ratio, and $x \propto Z_v$ denotes executing visual re-injection from $x$ to visual features $Z_v$. This operation re-injects visual evidence into the intermediate layer without introducing additional trainable parameters, significantly enhancing the model's faithfulness to the input visual content.

Another challenge comes from the similarity scores generated by MLLMs. We empirically find that multiple candidates may receive the same semantic score in Eq.~\ref{eq2}, leading to ambiguity in ranking. To resolve such ties, we introduce an entropy-based confidence calibration strategy. Specifically, we design a confidence-aware instruction to measure the model uncertainty of the query and candidate $(q,c)$ pair: ``\textit{\textless query\textgreater}, \textit{\textless candidate\textgreater}. Does the candidate match the query, True or False.". Subsequently, the uncertainty score is obtained as the normalized entropy of the output logits at the last token:
\begin{equation}
    H_i = -\sum_{v=1}^V p_v \log p_v,
\end{equation}
where $V$ is the vocabulary size, and $p_v$ is the softmax probability of token $v$ in the model's output distribution. Lower entropy $H_i$ indicates higher model certainty.
Among candidates that share the same semantic score, we select the one with the minimum entropy:
\begin{equation}
    C^* = \arg\min_{C_i \in \mathcal{P}} H_i,
\end{equation}
where $\mathcal{P}$ denotes the set of candidates with the same top-1 score. This confidence-aware selection strategy helps to refine the ranking when semantic distinctions are subtle, improving the reliability of the final retrieval results.
\section{EXPERIMENTS}
\label{sec:pagestyle}
\subsection{Datasets and Baselines}
To comprehensively assess the effectiveness of our proposed \texttt{RetLLM}, we evaluate it in a zero-shot setting on six benchmarks: Flickr30K~\cite{young2014image}, COCO~\cite{lin2014microsoft}, ShareGPT4V~\cite{chen2024sharegpt4v}, Urban1K~\cite{zhang2024long}, SugarCrepe~\cite{hsieh2023sugarcrepe}, and the MMEB~\cite{jiang2024vlm2vec} benchmark. For comparison, we include several strong baselines: CLIP, EVA-CLIP~\cite{sun2024eva}, E5-V~\cite{jiang2024e5}, VLM2Vec~\cite{jiang2024vlm2vec}, SigLIP~\cite{zhai2023sigmoid}, and UniME~\cite{gu2025breaking}, all evaluated under the same zero-shot protocol.
\subsection{Implementation Details and Evaluation Metrics}
Our main experiment employs Qwen2.5-VL-7B~\cite{Qwen2.5-VL} as the MLLM and CLIP-ViT-L/14@336px for coarse-stage retrieval, with top-5 candidates passed to the fine stage. Key enhancements include: 1) Visual re-injection ($\alpha$=0.3) to mitigate hallucination; 2) Entropy-based decision for ambiguous top scores. For ablation, we vary CLIP backbones (ViT-B, Long-CLIP-L\cite{zhang2024long}) and MLLMs (Phi-3.5-V~\cite{abdin2024phi3technicalreporthighly}, Qwen2-VL~\cite{Qwen2-VL}). All experiments are zero-shot. Primary metrics: Recall@1, measuring queries with correct target ranked first. On MMEB, we report average Precision@1 over all meta-tasks.
\begin{table*}[th]
\centering
\vspace{-1mm}
\setlength\tabcolsep{7pt}
\renewcommand\arraystretch{1}
\small
\fontsize{7pt}{7pt}\selectfont
\resizebox{\textwidth}{!}{
\begin{tabular}{lcccccccc}
\toprule
\multicolumn{1}{l}{{\textbf{Models}}} & \multicolumn{1}{c}{{\textbf{\#Parameters}}} & \multicolumn{4}{c}{\textbf{Per Meta-Task Score}} & \multicolumn{3}{c}{\textbf{Average Score}}    \\ \cmidrule(lr){3-6} \cmidrule(lr){7-9} 
\multicolumn{1}{c}{} & \multicolumn{1}{c}{} &  Classification & VQA & Retrieval & Grounding & IND & OOD & Overall \\ \midrule
\# of Datasets $\rightarrow$  & & 10       & 10            & 12            & 4             & 20            & 16            & 36            \\ \midrule
CLIP(ViT-L) & 0.4B & 42.8 & 9.1 & 53.0 & 51.8 & 37.1 & 38.7 & 39.2 \\
OpenCLIP(ViT-L) &  0.4B & 41.5 & 6.9 & 44.6 & 53.5 & 32.8 & 36.0 & 36.6 \\
SigLIP(So/14) & 0.9B & 40.3 & 8.4 & 31.6 & 59.5 & 32.3 & 38.0 & 35.0 \\
CLIP(ViT-BigG/14) & 2.5B & 52.3 & 14.0 & 50.5 & 60.3 & 38.9 & 45.8 & 44.3 \\
EVA-CLIP & 8B &  56.0 & 10.4 & 49.2 & 58.9 & 38.1 & 45.6 & 43.7   \\
E5-V & 7B & 39.7 & 10.8 & 39.4 & 60.2 & 34.2 & 33.4 & 37.5 \\
UniME & 7B & 43.0 & 17.7 & 42.5 & \textbf{63.2} & 37.6 & 38.6 & 41.6\\ 
\midrule
\rowcolor[HTML]{EDEDED}
\texttt{RetLLM} & 7B & \textbf{60.3} & \textbf{27.8} & \textbf{62.4} & 60.2 & \textbf{52.0} & \textbf{50.2} & \textbf{54.2} \\ 
\midrule
\bottomrule
\end{tabular}
}
\caption{Results on the MMEB benchmark. IND represents the in-distribution dataset, and OOD represents the out-of-distribution dataset. The reported scores are the average Precision@1 over the corresponding datasets on \textbf{zero-shot} manner. The best results are marked in bold.}
\label{tab:mmeb}
\end{table*}
\subsection{Results Analysis}
\texttt{RetLLM} achieves strong zero-shot performance across all benchmarks without any training or fine-tuning. As shown in Table~\ref{tab:diverse}, it consistently outperforms both zero-shot baselines (e.g., CLIP, EVA-CLIP) and MLLM retrievers such as E5-V and VLM2Vec. For example, on Flickr30K ($q^i \to c^t$), \texttt{RetLLM} reaches 94.5\% R@1, surpassing E5-V (88.7\%) and VLM2Vec (90.6\%). On ShareGPT4V ($q^t \to c^i$), \texttt{RetLLM} achieves 94.2\% R@1, outperforming VLM2Vec (86.9\%). On SugarCrepe ``Add'', \texttt{RetLLM} achieves 96.2\%, a 2\% gain over VLM2Vec (94.2\%), demonstrating superior zero-shot reasoning. On MMEB (Table~\ref{tab:mmeb}), \texttt{RetLLM} obtains 54.2\% overall Precision@1, a 12.6\% improvement over the strongest zero-shot baseline, UniME. It excels in Retrieval (62.4\%), Classification (60.2\%), and VQA (27.8\%), proving its robustness in zero-shot scenarios. These results confirm that, with our coarse-then-fine pipeline, visual enhancement, and entropy-based selection, \texttt{RetLLM} serves as a powerful zero-shot retriever.
\begin{table}[h]
\centering
\setlength\tabcolsep{3pt}
\renewcommand{\arraystretch}{0.9}
\begin{tabular}{lccccc}
\toprule
\textbf{Components} & \multicolumn{2}{c}{\textbf{Flickr30k}} & \multicolumn{2}{c}{\textbf{COCO}} \\
\cmidrule(lr){2-3} \cmidrule(lr){4-5}
 & $q^i \to c^t$ & $q^t \to c^i$ & $q^i \to c^t$ & $q^t \to c^i$ \\
\midrule
ALL & 94.5 & 81.8 & 69.2 & 52.1 \\
entropy only & 94.0 & 81.2 & 68.7 & 50.8 \\
enhancement only & 94.0 & 80.7 & 68.8 & 51.5 \\
MLLM only & 93.6 & 80.2 & 66.9 & 50.3 \\
\bottomrule
\end{tabular}
\caption{Ablation study of visual enhancement and entropy-based selection on Flickr30k and COCO.}
\label{tab:performance_comparison}
\end{table}
\vspace{-\baselineskip}
\subsection{Ablation Study}

\paragraph*{Components Effectiveness} 
As shown in Table~\ref{tab:performance_comparison}, removing visual enhancement causes a notable 1.5\% drop on COCO ($q^i \to c^t$), confirming its critical role in preserving visual fidelity during zero-shot retrieval. Disabling entropy-based selection leads to a 1.1\% decrease on Flickr30K ($q^t \to c^i$), demonstrating its effectiveness in resolving ambiguous rankings. The consistent superiority of the full model over ``MLLM only'' underscores the synergistic gain from combining both components.

\begin{table}[h]
\centering
\setlength\tabcolsep{3pt}
\renewcommand{\arraystretch}{0.8}
\begin{tabular}{lcccc}
\toprule
\textbf{CLIP-version} & \multicolumn{2}{c}{\textbf{ShareGPT4V}} & \multicolumn{2}{c}{\textbf{Urban1k}} \\
\cmidrule(lr){2-3} \cmidrule(lr){4-5}
 & $q^i \to c^t$ & $q^t \to c^i$ & $q^i \to c^t$ & $q^t \to c^i$ \\
\midrule
CLIP-ViT-B & 94.8 & 88.1 & 84.0 & 71.8 \\
CLIP-ViT-L & 97.6 & 94.2 & 88.9 & 78.6 \\
Long-CLIP-L & 96.6 & 95.1 & 95.2 & 95.8 \\
\bottomrule
\end{tabular}
\caption{Performance comparison using different CLIP versions with fixed Qwen2.5-VL.}
\label{tab:clip_comparison}
\end{table}

\begin{figure}[!th]
\centering
\begin{adjustbox}{width=0.48\textwidth}
\includegraphics{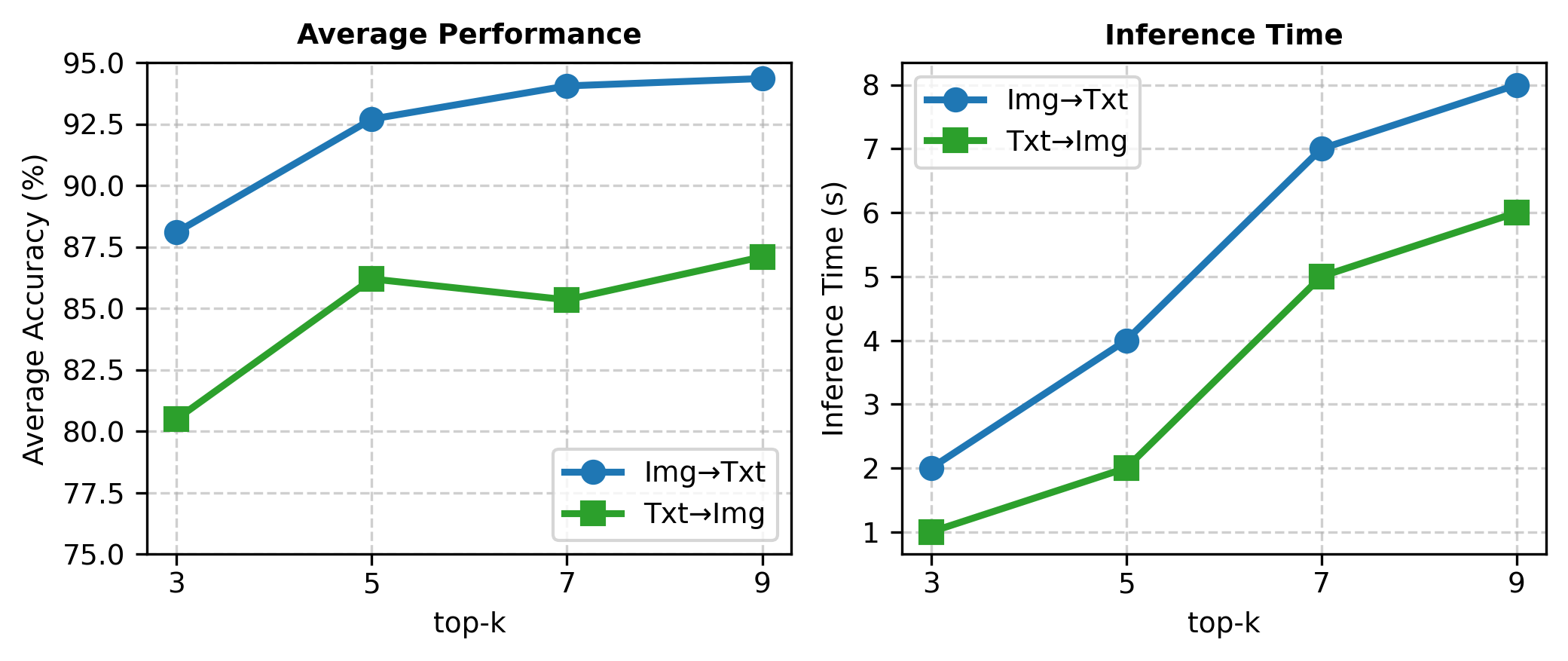}
\end{adjustbox}
\caption{Ablation studies on the impact of top-k values on retrieval performance and inference efficiency.}
\label{abla2}
\end{figure}

\vspace{-\baselineskip} 
\paragraph*{Top-$k$ Sensitivity} As shown in Fig.~\ref{abla2}, performance varies with different $k$ values (3, 5, 7, 9), revealing a clear trade-off between precision and efficiency: a larger $k$ improves recall at higher computational cost, while $k$=5 (our default) offers the optimal balance for practical deployment.
\begin{table}[h]
\centering
\setlength\tabcolsep{5pt}
\renewcommand{\arraystretch}{0.9}
\begin{tabular}{lcccc}
\toprule
\textbf{MLLMs} & \multicolumn{2}{c}{\textbf{ShareGPT4V}} & \multicolumn{2}{c}{\textbf{Urban1k}} \\
\cmidrule(lr){2-3} \cmidrule(lr){4-5}
 & $q^i \to c^t$ & $q^t \to c^i$ & $q^i \to c^t$ & $q^t \to c^i$ \\
\midrule
phi3.5v & 86.5 & 72.3 & 78.9 & 73.5 \\
Qwen2-VL & 93.8 & 94.1 & 87.2 & 78.2 \\
Qwen2.5-VL & 97.6 & 94.2 & 88.9 & 78.6 \\
\bottomrule
\end{tabular}
\caption{Performance comparison using different MLLMs with fixed CLIP-ViT-L.}
\label{tab:MLLM_comparison}
\end{table}

\vspace{-\baselineskip} 
\paragraph*{Model Scalability}
As shown in Table~\ref{tab:clip_comparison} and Table~\ref{tab:MLLM_comparison}, RetLLM benefits from stronger and larger backbone models, with performance improving consistently as the capacity of the underlying components increases. This highlights the scalability of our framework and its ability to leverage advances in both CLIP models and multimodal large language models.
\section{CONCLUSION}
\label{sec:conclusion}
In this work, we propose \texttt{RetLLM}, a training-free multimodal retrieval framework that achieves strong zero-shot performance through coarse-then-fine search, visual enhancement, and entropy-based selection. Crucially, \texttt{RetLLM} is highly scalable: it naturally inherits performance gains from stronger foundation models in a plug-and-play manner, making it a forward-compatible and sustainable solution for future retrieval systems.
\clearpage
\footnotesize
\bibliographystyle{IEEEbib}
\bibliography{strings,refs}

\end{document}